
\magnification=1200

\vsize = 24truecm
\hsize = 16truecm

\null
\vskip-1truecm
\rightline{SISSA Ref. 159/95/EP}
\rightline{PREPRINT}
\vskip1truecm
\medskip
\vskip2truecm
\centerline{{\bf THE LAX PAIR BY DIMENSIONAL REDUCTION }}
\bigskip
\centerline  {{\bf
OF CHERN-SIMONS GAUGE THEORY}}
\vskip1.5cm
\centerline{ O.K.Pashaev
\footnote{$^*$}{Permanent address: Joint Institute for
Nuclear Research,  Dubna 141980, Russian Federation. E-mail:
pashaev@main1.jinr.dubna.su} }
\medskip
\centerline{\it International School for Advanced Studies (SISSA/ISAS)}
\centerline{\it Via Beirut 2, 34014 Trieste}
\centerline{\it Italy.}
\vskip1cm
\centerline{ABSTRACT}
\baselineskip=16pt
\medskip
We show that the Nonlinear Schr\"odinger Equation and
the related Lax pair
in 1+1 dimensions can be derived
from  2+1 dimensional Chern-Simons Topological Gauge Theory.
The spectral parameter, a  main object for the Loop algebra structure
and the Inverse Spectral Transform, has
appear as a
homogeneous part (condensate) of the
statistical gauge field, connected with the compactified extra space
coordinate. In terms of solitons, a  natural interpretation for the
one-dimensional analog of Chern-Simons Gauss law
is given.

\vskip1cm
\centerline{November 1995}

\vfill\eject

\noindent
{\bf 0 Introduction}
\par \medskip
As well known, it was Kaluza and Klein idea for unification of gravitation and
electromagnetism in a 5 dimensional theory of gravity [1]. Then, the idea was
developed in the context of dual models and the string theory, consistently
quantized  only in
10 and 26 space-time dimensions [2]. The significant attempt
in the higher dimensional interpretation of internal symmetries was connected
with a special kind of the
compact space like the Calabi- Yau space. Hence, the
ordinary 4-dimensional
physics, completely defined by the massless modes,
turns out to be described as
a low energy approximation of a bigger theory.
\par
An important aspect of the problem consists  an intimate relation between
the dimensional reduction procedure and the
nonlinear models. In fact, if the linear modes in the theory always
trivially
decouple, the dimensional reduction of nonlinear models,
in general, is not trivial.
Thus, the Yang-Mills theory in 4+N dimensions leads, via dimensional reduction,
to a Yang-Mills + Higgs scalars coupled theory with specific couplings
[3].
\par
In this
context we can imagine that the integrability of some nonlinear
models can be
related to  the
dimensional reduction procedure.
This guess is indicated by a "folk theorem" that dimensional reduction from
higher dimensions enlarges the symmetry $G$ to its affine extension
[4][5].
Then,
some infinite-dimensional symmetries, appearing as the hidden symmetries of
integrable  models, shall have a geometrical meaning.
\par
Moreover, by dimensional reduction,
many 0+1 and 1+1 dimensional models
were embedded
to the self-dual Yang-Mills (SDYM) equations [6]. These equations are
an integrable system
admitting the linear representation, or  the Lax pair [7].
By suitable reduction,  the Lax pair associated with corresponding
low dimensional model has
appeared from the Lax pair for SDYM.
Furthermore, one believes even that the self-dual Yang-Mills equations
are a universal
integrable system from which all the others could be obtained by proper
reductions [6]. This
programme,
still being intensively studied,
is closely connected with the twistor description and
requires that there should be a linear system for equations of the
zero-curvature type (the Lax pair). However, the origin of the linear
system remains
a {\it terra incognita}. As well, as the most mysterious  part of the linear
problem
- the spectral
parameter. A time independence of the spectral parameter usually
 is connected
with an
infinite number
of integrals of motion, while the integrable dynamics is produced by the
isospectral deformations. From the algebraic point of view, the presence
of spectral
parameter in the linear problem with Lie algebra ${\cal G}$,
announces the appearance of enlarged,
loop algebra structure  ${\cal G} \times C[\lambda,\lambda^{-1}] $,
associated with  hidden non-Abelian symmetry of the model.
\par Another important point is
that the spectral parameter is present in the linear problem
and absent in the related evolution equation. Since the last one arises from
zero-curvature condition (ZCC) for the associated flat connections,
it suggests a
 gauge-theoretical formulation
of this phenomena. According to this observation we expect the existence of
non-Abelian gauge theory, which includes the spectral parameter as a gauge
degree of
freedom. Hence, the isospectral deformation defined by
nonlinear evolution equation
should
appear as a gauge invariant condition.
\par Thus, we are looking for non-Abelian gauge
theory with symmetries not
less than integrable one. That means the existence of an
infinite number of integrals of motion and related hierarchy of different
time evolutions. Usually,
in the high energy physics,
the  unification procedure
means an embedding to a larger symmetry group. Naturally, we can suppose  that
the
unified model could be some type of the Conformal Field Theory.
In fact, some integrable models after proper limiting procedure show
conformal properties (Faddeev's approach)[8]. In opposite way,
a proper breaking of the
conformal invariance  leads to an  integrable model (Zamolodchikov's
approach)[9].
Thus, the existence of an  infinite number of conservation laws is some relict
of
conformal symmetry.
However, as we know, a different type of integrability property
exists: $C -$ and $S$ integrability, Darboux
 integrability [10].
And some of them  are very
strict. Actually, the Liouville equation is conformal invariant, Darboux
 integrable and
C - integrable. But the affine Liouville [11] and the nonlinear $\sigma-$
model,
being conformal invariant, seems as Darboux non-integrable [12].
Moreover, various integrable models also in three and four dimensions exist.
\par
Apparently most drastic possibilities for unification
provide  the Topological
Field Theory (TFT) [13]. As well known [14], TFT admits a huge
diffeomorphism symmetry, which realized by gauge transformations. Resulting
 reparametrization
invariance of the
model leads to the trivial
dynamics, frozen in the reparametrization of gauge
(unphysical) parameters.
\par As it was shown,
the dimensional reduction idea is very useful in the TFT [15].
Thus, the 3-dimensional gauge field theoretical
formulation of TFT in the Chern-Simons (CS) form [16]
can be dimensionally reduced from the 4 -dimensional TFT,
which close relates to the self-dual Yang-Mills instantons and the
Donaldson theory. Moreover, by dimensional reduction, the   conformal field
theory
in 2- dimensions
was obtained [17].
A general
 reduction of CS theory
 leads to 2-dimensional TFT, known as the BF theory [18].
Peculiar property of
2-dimensional
BF theory is that equations of motion have a zero-curvature form.
Conformal field
theory in the Liouville form
and
the  Toda theory has  been related also to BF
theory [19]. Futhermore, the two linear gravities - based on the de Sitter
group or a central extension of the Poincare group - were
derived from 3-dimensional TFT [20].
\par
These results suggest that TFT could be a good candidate for the universal
model,
properly reducing of which conformal invariant and integrable systems
can be obtained.
The main problem arises how to break topological symmetries. If we start with
2- dimensional BF
theory, obviously we can reduce field connections to the form, related to
specific integrable model. But it means, that we need to put "by hand"
the spectral
parameter dependence  of connections. Then, the gauge
group for the BF theory should be some kind of the loop group. Hence,
 the loop structure of the problem
suggests that we can try to do dimensional reduction of the model starting
from 3-dimensional
CS theory. The question now only is how constraint the model to have
an integrable
2-dimensional system.
\par
{}From another hand, for
nonlinear $\sigma$ models some constraint equations naturally arise.
The idea, inspired by the gauge relation between one dimensional
integrable models, is to use variables from the tangent space to the nonlinear
manifold [21].
By this approach, some evolution $\sigma-$ models like the
Heisenberg Model (HM) and
the Topological Magnet, are reformulated     as the
U(1)  gauge invariant field theory [22-24].
A mapping of the model to 3-dimensional zero-curvature condition
(or
to the CS theory)
implies that the field connection
 should satisfy to proper constraint.
In contrast with time reparametrization invariance of CS theory,  the reduced
system evolves according to related $\sigma$ model. For integrable
evolution [23] it means a breaking of continual, time reparametrization
symmetry of TFT up to discrete time hierarchy of integrable models.
\par
In the present paper we show that 2+1 dimensional HM, considered
as a constraint for CS theory, provides
by dimensional reduction not only integrable model,
the  Nonlinear Schr\"odinger Equation (NLSE),
 but also the corresponding Lax
pair. The spectral parameter appears automatically in a correct way
and has the meaning of  homogeneous (condensate) part for the statistical
gauge
field, related with the extra space dimension.
Moreover, non-homogeneous structure of the field
is related to the B\"acklund transformations for NLSE.\par
We speculate that
situation could have the general meaning and is applicable to other
integrable models.
In fact, all what we need are constraints for CS theory, arising from
nonlinear $\sigma$ model.
Then, after dimensional reduction,
 an integrable model and the related Lax pair with proper
graded structure should appear automatically.
Our result suggests   that while the Lax pair depends on additional
spectral parameter, which is remnant of the three-dimensional theory,
the reduced Nonlinear Evolution Equation is  gauge invariant and
independs of that parameter. Apparently this idea helps  to clarify
 another well known "folk theorem": a standard way to construct integrable
models
in 2+1 dimensions is to start with the Lax pair for a 1+1 dimensional
integrable model, and then replace the spectral parameter by a differential
operator in extra direction. For example, this gives the Lax pair for the
Kadomtsev-
Petviashvili (KP) equation from the linear system for the KdV equation.
The theorem indicates to  another interpretaion of the spectral parameter
as the canonically conjugate (momentum) variable to
the extra space coordinate. It seems that after quantization the relation
between
two ideas will be more transparent and
 1+1 and 2+1 dimensional integrable models
could be related to different (coordinate - momentum) representations
of TFT.  Worthe to note also on possibility to track connection between
quantum exactly soluble TFT and integrable models in the spirit of [25].
\par Perhaps most ambitious programm is to obtain the
Lax pair for the SDYM from TFT in higher then 4 dimensions.
However,
 up to
now only known TFT that can be defined in arbitrary dimensions is BF theory.
\par In Sec.I, we present the general formalism of constructing the gauge
invariant field theory, associated with the nonlinear $\sigma$ model.
Sec.II describes the related formulation of the non-Abelian CS theory.
In Sec.III, we illustrait the general formalism with two important
examples. Dimensional reduction for 2+1 HM will be considered in
Sect.IV. In Conclusion we discuss some physical ideas to explain our results.
The one-dimensional analog of CS Gauss law we interpret in terms of
solitons for integrable models.
\bigskip
\medskip
\noindent {\bf 1.Adjoint Representation of ZCC }
\par
\bigskip
\par
In this section we present  a  general  formalism  connecting  a  zero
curvature equations on $A_{1}$ algebra $(SU(2)$ or any non-compact version of
it) in the adjoint representation with moving trihedral [26]. This
formalism allows us formulate a nonlinear $\sigma $  model as the Abelian gauge
field theory.
\par
Let us consider the group $A_{1}$ with element $g$, generated by
$\tau_{i} (i=1,2,3)$ , satisfying
$$
\tau_{i}\tau_{j} = h_{ij} + ic_{ijk}\tau_{k},
\eqno(1.1)$$
where $h_{ij}$ and $c_{ijk}$ are the Killing metric
and structure constants of
$A_{1}$. We define an orthonormal trihedral set of unit vectors
${\bf n}_{i}$
and ${\bf e}_{i}$ , and matrices $N_{i}$ and $E_{i}$
correspondingly, in the adjoint
representation
$$
N_{i} = ({\bf n}_{i},\tau) =
{\bf n}^{k}_{i}\tau_{k} = h_{kl}{\bf n}^{k}_{i}\tau^{l} =
g \tau_{i} g^{-1},\eqno (1.2a)$$
$$
E_{i} = ({\bf e}_{i},\tau) = {\bf e}^{k}_{i}\tau_{k} =
h_{kl}{\bf e}^{k}\tau^{l} = g^{-1}\tau_{i}g .\eqno(1.2b)$$
Using (1.1), the orthonormality of the trihedral are expressed by relations
\par
$$
N_{i}N_{j} = h_{ij} + ic_{ijk} N_{k} ,\eqno (1.3a)$$
$$
E_{i}E_{j} = h_{ij} + ic_{ijk} E_{k} .\eqno (1.3b)$$
The Killing metric $h_{ij}$ and structure constants
$c_{ijk} = - c_{jik}$ defines the inner and
cross products between
three-vectors, transforming in the adjoint representation of $A_{1}$:
$$
({\bf n}_{i}{\bf n}_{j}) = h_{ij} ,\eqno (1.4a)$$
$$
{\bf n}_{i}\wedge  {\bf n}_{j} = c_{ijk} {\bf n}_{k},\eqno (1.4b)$$
(and the similar equations for ${\bf e}_{i}$ vectors).
Matrices $N_{i}$  and $E_{i}$ are
connected by the similarity transformation
$$
N_{i}= g^{2}E_{i} g^{-2},
\eqno(1.5)$$
while related ${\bf n}_{i}$ and ${\bf e}_{i}$ vectors satisfy
$$
({\bf n}_{i})^{j}h_{jj}= ({\bf e}_{j})^{i}h_{ii} ,
\eqno(1.6)$$ (no summation).
Due to this relation, in the present paper we restrict ourselves only with
${\bf n}_{i}$
vectors.
\par
Let ${\bf n}_{i}= {\bf n}_{i}(x)$ are a smooth
vector fields that define at each space coordinate ${\bf x}
= (x^{1},x^{2},x^{3})$ of
$ M$ , the three vectors $({\bf n}_{1}(x),{\bf n}_{2}(x),{\bf n}_{3}(x))$
forming an orthonormal
basis called the moving frame.
\par
We can introduce the left- and right-invariant chiral currents
$$
J^{R}_{\mu } = g^{-1}\partial _{\mu }g,  \eqno(1.7a)$$
$$
J^{L}_{\mu } = \partial _{\mu }g g^{-1},\eqno(1.7b)$$ $(\mu = 1,2,3)$.
They are connected by simple transformation
$$
J^{R}_{\mu } = g^{-1}J^{L}_{\mu } g .
\eqno(1.8)$$
The trihedral moves according to equations
$$
\partial _{\mu }N_{i}= [J^{L}_{\mu },N_{i}] = g[J^{R}_{\mu}
,\tau_{i}]g^{-1}, \eqno(1.9a)$$
$$
\partial _{\mu }E_{i}= g^{-1}[\tau_{i},J^{L}_{\mu }]g =
[E_{i},J^{R}_{\mu }] ,\eqno (1.9b)$$
or in the three-dimensional representation
$$
\partial _{\mu }N_{i}= (J^{R}_{\mu })_{ik}N_{k} ,\eqno(1.10a)$$
$$
\partial _{\mu }E_{i}= - (J^{L}_{\mu })_{ik}E_{k} ,\eqno (1.10b)$$
 where $(J^{R}_{\mu })_{ik}$ and $(J^{L}_{\mu })_{ik}$
 are matrices in the adjoint representation
$$
(J^{R,L}_{\mu })_{ik} = - ic_{ijk}(J^{R,L}_{\mu })_{j} =
 i(J^{R,L}_{\mu })_{j}c_{jik} ,
\eqno(1.11)$$
and $J^{R,L}_{\mu } = \sum  (J^{R,L}_{\mu })_{j} (1/2)\tau_{j}$ .
Related rotation of the moving frame is given by equations
$$
\partial _{\mu }{\bf n}_{i}= (J^{R}_{\mu })_{ik}{\bf n}_{k} ,
$$
$$\partial _{\mu }{\bf e}_{i}= - (J^{L}_{\mu })_{ik}{\bf e}_{k}.$$
\par
 Matrices $J^{R,L}_{\mu }$ have the symmetry property
$$
(J^{R,L}_{\mu })_{ij} h_{jj} = - (J^{R,L}_{\mu })_{ji} h_{ii} .
$$
For $SU(2)$ case $h_{ij} = \delta _{ij}$ , $c_{ijk} =
\epsilon _{ijk}$ and the matrices (1.11) are
anisymmetric. \par
The zero-curvature conditions for chiral currents (1.7) have the
 form
$$
\partial _{\mu }J^{R}_{\nu } -
\partial _{\nu }J^{R}_{\mu } + [J^{R}_{\mu }, J^{R}_{\nu }] = 0 ,
\eqno(1.12a)$$
$$
\partial _{\mu }J^{L}_{\nu } - \partial _{\nu }J^{L}_{\mu } -
[J^{L}_{\mu }, J^{L}_{\nu }] = 0 . \eqno(1.12b)$$
In the following discussion we concerned mainly on the $J^{R}_{\mu }$ matrix
and  skip the $R$ - index.
\par
Let us decompose  matrix $J_{\mu }$ to the diagonal and off diagonal
parts
$$
J_{\mu } = J^{(0)}_{\mu }+ J^{(1)}_{\mu },
$$
parametrized in the form
$$
J^{(0)}_{\mu } = i/4 \sigma _{3}V_{\mu } ,\eqno(1.13a)$$
$$J^{(1)}_{\mu } =
\left(\matrix{0&-\kappa^{2}\bar q_{\mu} \cr q_{\mu}&0 \cr}\right),
\eqno(1.13b)$$
where $\kappa ^{2}= + 1$ for $SU(2)$ and $\kappa ^{2}= - 1$
for $SU(1,1)$case. Then,  in  the
adjoint representation we have
$$
(J_{\mu })^{ad} = {1\over 2}
\left(\matrix{
0&V_{\mu}&4Re(q_{\mu})\cr -V_{\mu}&0&4Im(q_{\mu}) \cr
-4Re(q_{\mu})&-4Im(q_{\mu})&0 \cr}\right).
\eqno(1.14)$$
\par The moving frame rotates with $x$ variation according to equations
$$
\partial _{\mu }{\bf n}_{1}= -1/2 V_{\mu }{\bf n}_{2} -
2 (Re q_{\mu }){\bf n}_{3},
\eqno(1.15a)$$
$$
\partial _{\mu }{\bf n}_{2}= 1/2 V_{\mu }{\bf n}_{1} -
2 (Im q_{\mu }){\bf n}_{3},\eqno(1.15b)$$
$$
\partial _{\mu }{\bf n}_{3}= 2 (Re q_{\mu }) {\bf n}_{1} +
2 (Im q_{\mu }){\bf n}_{2}.\eqno(1.15c)$$
\par If we denote $U_{\mu } \equiv  (Re (q_{\mu}),Im (q_{\mu }))$,
the system can be
written in a more compact form
$$
\partial _{\mu }{\bf n}_{i}= -1/2 V_{\mu }
\epsilon _{ij}{\bf n}_{j} - 2 U_{i\mu }{\bf s} ,\eqno(1.16a)
$$
$$
\partial _{\mu }{\bf s} = 2 U_{i\mu }{\bf n}_{i} .
\eqno(1.16b)$$
\par The vector ${\bf s \equiv  n}_{3}$ satisfies the constraint
$$
({\bf s}(x), {\bf s}(x)) = h_{33} ,
\eqno(1.17)$$
where $h_{33}= 1$ for $SU(2)$ and $SU(1,1) $,
 (and $h_{33}= -1$ for $SL(2,R))$.
It belongs to the two-dimensional sphere $S^{2}$ or pseudosphere $S^{1,1}$
correspondingly.
\par
Fields $V_{\mu }$ and $q_{\mu }$ are given by projections
$$
V_{\mu } = -2({\bf n}_{2}, \partial _{\mu }{\bf n}_{1}), Re (q_{\mu }) =
-1/2({\bf s, \partial }_{\mu }{\bf n}_{1}), Im (q_{\mu }) =
-1/2({\bf s, \partial }_{\mu }{\bf n}_{2}).
\eqno(1.18)$$
Two vector fields $({\bf n}_{1}(x),{\bf n}_{2}(x))$ at each ${\bf x}$ form a
basis in the tangent space  to  corresponding  manifold  for ${\bf s}(x)$.
But vectors ${\bf n}_{1}$ and ${\bf n}_{2}$ are not
uniquely  determined  by  eq.(1.4).
\par
If  we choose other ${\bf n}_{1}' , {\bf n}_{2}'$ as a
rotated basis
$${\bf n}_{1}'= \cos  \alpha  {\bf n}_{1} -
\sin  \alpha  {\bf n}_{2} , {\bf n}_{2}'= \cos
\alpha  {\bf n}_{2} + \sin  \alpha  {\bf n}_{1} , \eqno(1.18)$$
related fields $V_{\mu }'$ and $q_{\mu }'$ defined by (1.18)
are the $U(1)$ gauge
transformed fields
$$
V_{\mu }'= V_{\mu } + 2 \partial _{\mu }\alpha  ,
q_{\mu }'= e^{i\alpha }q_{\mu }
\eqno(1.19)$$
Expression for $q_{\mu }$ field simplifies if we introduce a complex basis
$$
{\bf n}_{+}= {\bf n}_{1} + i {\bf n}_{2} , {\bf n}_{-}=
{\bf n}_{1} - i {\bf n}_{2} ,
\eqno(1.20)$$
 satisfying to following relations
$$
({\bf n}_{+}, {\bf n}_{+}) = 0 = ({\bf n}_{-}, {\bf n}_{-}),\eqno (1.20a)
$$
$$
({\bf n}_{+}, {\bf n}_{-}) = 2,\eqno(1.20b)
$$
$$
{\bf n}_{+}\times  {\bf s} = i{\bf n}_{+} ,
{\bf n}_{-}\times  {\bf s} = -i{\bf n}_{-} ,
{\bf n}_{-}\times  {\bf n}_{+} = 2 i {\bf s}. \eqno(1.20c)
$$
Then
$$
q_{\mu } = 1/2 (\partial _{\mu }{\bf s}, {\bf n}_{+}), \bar{q}_{\mu } =
 1/2 (\partial _{\mu }{\bf s}, {\bf n}_{-}) .
\eqno(1.21)$$
 In terms of (1.20) the moving frame equations (1.16) become
$$
D_{\mu }{\bf n}_{+} = -2 q_{\mu } {\bf s} ,\eqno(1.22a)
$$
$$
\partial _{\mu }{\bf s} = q_{\mu } {\bf n}_{-} +
\bar{q}_{\mu } {\bf n}_{+} ,\eqno (1.22b)
$$
where $D_{\mu } \equiv  \partial _{\mu } - i/2 V_{\mu }$
is the covariant derivative.
\par
This   form   is   explicitly invariant under the local $U(1)$
gauge transformations
$$
{\bf s} \rightarrow {\bf s} , {\bf n}_{+}\rightarrow e^{i\alpha }
{\bf n}_{+} , {\bf n}_{-} \rightarrow e^{-i\alpha } {\bf n}_{-} ,
\eqno(1.23)$$
 which are just the local rotations in the tangent to the
vector {\bf s} plane.
\par
As follows from eqs.(1.22) fields $V_{\mu }$ and $q_{\mu }$ are subject
to the system
$$
D_{\mu }q_{\nu }= D_{\mu }q_{\mu }\qquad , \eqno(1.24a)
$$
$$
[D_{\mu },D_{\nu }] = -2(\bar{q}_{\mu }q_{\nu } -
\bar{q}_{\nu }q_{\mu }) . \eqno(1.24b)
$$
\par To describe a time evolution of three-dimensional physical system we
 need introduce the space-time $M_{3}= T \times  M_{2}$ decomposition,
where $T$ is associate with the time variable $x_{3}= t$ and $M_{2}$ is a 2 -
dimensional space manifold. In this  case  a  time  evolution  of  the
moving frame
$$
D_{0}{\bf n}_{+} = -2 q_{0} {\bf s} ,\eqno(1.25a)
$$
$$\partial _{0}{\bf s} = q_{0} {\bf n}_{-} + \bar{q}_{0}
{\bf n}_{+} ,\eqno(1.25b)$$
is  completely an arbitrary due to
arbitrarines of $q_{0}$. We recall
that $q_{0}$ as well as $V_{0}$ are Lagrange multipliers of the CS TFT
appearing in front of the CS Gauss law of the theory (see eq.(2.12)).
Moreover, Eq. $(1.25b)$
shows  that
evolution  of  the  spin  vector ${\bf s}$  associated  with CS TFT
, being $U(1)$ gauge invariant, remains
completely  an  arbitrary. In  this  sense the  TFT  are  related  to   the
nonlinear $\sigma $  model  with  an arbitrary   evolution   (reparametrization
invariance) or, what is the same, without any evolution, modulo $U(1)$ gauge
transformations.
\par
Formally we can represent eq.$(1.25b)$ as the spin precession equation
$$
\partial _{0}{\bf s} = {\bf s \times  H}
\eqno(1.26)$$
in arbitrary $U(1)$ gauge invariant magnetic field
$$
{\bf H} = i(\bar{q}_{0} {\bf n}_{+} - q_{0} {\bf n}_{-})
\eqno(1.27)$$
\par
 Nevertheless to above  results a topological
restriction
on the  possible  spin configurations exists. Indeed,  we  can  imagine  that
the space $M_{2}$ is compact. For example if we suppose that the value of
the spin vector ${\bf s}$ at infinity is fixed ${\bf s} \rightarrow (0,0,1)$.
Then,  all  smooth
configurations describing mapping of $(x_{1}, x_{2})$ into ${\bf s}(x)$,
independently of the evolution, are
classified by  the integer valued degree of mapping of $S^{2}\rightarrow
 S^{2}$, or the topological charge:
$$
Q = {1\over{8\pi}}\int \epsilon _{ij}{\bf s}(\partial _{i}{\bf s \times
\partial }_{j}{\bf s})d^{2}x = {1\over{8\pi}}
\int\epsilon _{ij}\partial _{i}V_{j} d^{2}x
.\eqno(1.28)$$
\par In terms of our gauge fields, the topological charge
density has the form
$$
\epsilon _{ij}{\bf s}(\partial _{i}{\bf s \times
\partial }_{j}{\bf s}) = \epsilon _{ij}\partial _{i}V_{j} = B,
\eqno(1.29)$$
 of the radial (along the ${\bf s})$ oriented magnetic
 field B associated with the
vector potential $V_{j}$. As well known, eq.(1.28) states that the winding
number of mapping $S^{2}\rightarrow S^{2}$  coincides
with the winding number of the
 mapping of the circle $S^{1}$ at $x^{2}_{1} +
 x^{2}_{2} \rightarrow \infty $ into the abelian gauge group
 manifold. It means that all $U(1)$ gauge transformations (1.19)
also fall into topological classes characterized by winding number
(1.28). Just substituting (1.19) in to (1.28) we find that under
Abelian gauge transformations
$$
V_{j} \rightarrow V_{j} + 2\partial _{j}\alpha ,
\eqno(1.30)$$
$Q$ transforms as
$$
Q \rightarrow Q + {1\over{4\pi }}\int \epsilon _{ij}
\partial _{i}\partial _{j}\alpha d^{2}x .
\eqno(1.31)$$
For a smooth gauge transformation the second term vanishes
and $Q$ is invariant.
\par
 More generally, if $M_{2}$ is a compact Riemann surface of genus $g$ ,
$M_{2} = \Sigma _{g} $, the charge $Q$ in (1.28)
is  the first Chern class $c_{1}$ , which is an
integer [27].
\par
However, if $M_{2}$ admits some singular points, $Q$ could be  an arbitrary
number. Let us consider a potential $V_{j}$ with charge $Q.$ We
perform a singular at ${\bf x} = 0$ rotation (1.31)
$({\bf n}_{1},{\bf n}_{2})$ with angle
$$
\alpha ({\bf x}) = N\theta ({\bf x}) ,
\eqno(1.32)$$
 where $\theta ({\bf x}) = \arctan  (x_{2} /x_{1} )$.
 Then, using unconventional representation
for the planar $\delta $-function [28]
$$
\epsilon _{ij} \partial _{i}\partial _{j}\theta  =
(\partial _{1}\partial _{2} - \partial _{2}\partial _{1})\theta  =
2\pi  \delta ^{2}(x),
\eqno(1.33)$$
we find that
$$
\Delta Q = N/2 .\eqno(1.33)
$$ \par
As  evident,  instead  of  integer $N$  we  can  use  an
arbitrary real number which gives us arbitrary Q.
This singular gauge  transformation is  related  with
a point vortex
creation
at ${\bf x} = 0$ and is described by an anyon potential
$$
V^{A}_{i} = 2 {\partial \over{\partial x_{i}}}\alpha ({\bf x}) =
2N {\partial \over{\partial x_{i}}}\theta ({\bf x}) =
$$ $$-2N
\epsilon _{ij}\partial _{j} \ln |{\bf x|}
= -2N \epsilon _{ij}{x_{j} \over{|{\bf x|}^{2}}}.
\eqno(1.34)$$\par
In a more  general situation, for $n$ point vortices  located  at
$
{\bf x}_{1} ,\ldots
,{\bf x}_{n}
$,
with related strength $N_{p} (p = 1,\ldots
,n)$, the  vector potential
$$V^{A}_{i} ({\bf x} ;{\bf x}_{1},\ldots
,{\bf x}_{n})
= -2 \epsilon _{ij} \sum^{n}_{p=1}N_{p}{ (x^{j} -
x^{j}_{p} )\over{|{\bf x} - {\bf x}_{p}|^{2}}} =
-2 \epsilon _{ij}\partial _{j} \sum^{n}_{p=1}N_{p} \ln |{\bf x} -
{\bf x}_{p}| ,
\eqno(1.35)$$
produces the vanishing almost everywhere magnetic field
$$
B({\bf x}) = \epsilon _{ij}\partial _{i}V_{j} =
4\pi  \sum^{n}_{p=1}N_{p} \delta ^{2}(x - x_{p} ) .
\eqno(1.36)$$
The corresponding charge changes as
$$
\Delta Q = 1/2\sum^{n}_{p=1}N_{p} .
\eqno(1.37)$$ \bigskip
\noindent {\bf 2.Chern-Simons Gauge Theory reduction}
\bigskip
\par
In previous section we introduced the chiral fields
$J_{\mu } (1.7)$ satisfying to the zero curvature condition (1.12). The last
one in term of  components (1.13) is described by the system
(1.24). For fields $V_{\mu }$ and $q_{\mu }$ , subject to  (1.24) ,
the moving frame  can  be  reconstructed  from  eq.(1.16).
Moreover,  the
current $J_{\mu }$  can be considered as  non-Abelian pure gauge
potential. Then the zero- curvature equations (1.12) are of the
Lagrangian form for pure non-Abelian Chern-Simons functional.
\par
The Cern-Simons action is defined as follows
$$
S[J] = {k\over{4\pi}} \int_{M} Tr(J\wedge dJ + {2\over 3} J\wedge J\wedge J),
\eqno(2.1)$$
 where $M$ is an  oriented  three-dimensional  manifold, $J$ is a
gauge connection with values in the  Lie  algebra  {\cal G}.  Action  (2.1)
is manifestly independent from the space metric, so it was interpreted
by Witten as a general covariant theory, or topological field theory
[1].
\par
The classical equations of motion following from action (2.1)  have
the form
$$
F = dJ + J\wedge J = 0
\eqno(2.2)$$
 of zero-curvature condition.
\par
To adopt the canonical approach to the problem, one considers a  region
of the three-manifold to be isomorphic to $M_{3}= T \times  M_{2}$, where we
interpret $T$
as the time. Then, for the gauge
field we have $J_{\mu }= (J_{0},J_{j})$, where $J_{0}$ is
the time component and the action (2.1) takes the form
$$
S(J) = - k/4\pi \int_{\Sigma}\int dt
\epsilon ^{ij} Tr(J_{i}{\partial\over{ \partial t}} J_{j} - J_{0} F_{ij}),
\eqno(2.3)$$
 where
$$
F_{ij}= \partial _{i}J_{j}- \partial _{j}J_{i} + [J_{i},J_{j}] .\eqno(2.3a)
$$
In the basis
\par
$$
T_{a}= {1\over2} \tau_{a} ,(a = 1,2,3),\eqno(2.4a)
$$
$$
[T_{a},T_{b}] = ic_{abc}T_{c} ,\eqno(2.4b)
$$
with $$Tr(T_{a}T_{b}) = {1\over 2} h_{ab},\eqno(2.4c)
$$
(see eq.(1.1))
we have the Poisson brackets
for components $J_{\mu } = \sum^{3}_{a=1} (J_{\mu })_{a}T_{a}$:
$$
\{J^{a}_{i}(x),J^{b}_{j}(y)\} =
{4\pi \over k}\epsilon _{ij}h^{ab}\delta ^{2}(x - y).
\eqno(2.5)$$
Then, in terms of $V_{\mu }$ and $q_{\mu }$ fields
$$
\{V_{i}(x),V_{j}(y)\} = - {16\pi \over k}\epsilon _{ij}h_{33}\delta ^{2}(x - y)
,
\eqno(2.6)$$
$$
\{Re (q_{i}(x)), Re (q_{j}(y))\} =
- {\pi \over k}\epsilon _{ij}h_{11}\delta ^{2}(x - y),\eqno(2.7a)
$$
$$
\{Im (q_{i}(x)), Im (q_{j}(y))\} = -
{\pi \over k}\epsilon _{ij}h_{22}\delta ^{2}(x - y).\eqno(2.7b)
$$
The last two relations have more appropriate form if we
introduce new fields,
(idea was inspired by the gauge relation between 1+ 1 dimensional
NLSE and HM),
$$
\psi _{\pm } = {1\over{2\sqrt{\pi}}}  (q_{1}\pm iq_{2}) .
\eqno(2.8)$$
They are directly related with the complex structure on the manifold $M_{2}$
in terms of
$$z = x_{1} + ix_{2}  , \bar z = x_{1} - ix_{2}. \eqno(2.9)$$
\par
The Poisson brackets for $\psi _{\pm }$ fields are
$$
\{\psi _{+}(x), \bar\psi_{+}(y)\} =
{i\over{2k}}(h_{11}+ h_{22})\delta ^{2}(x - y),\eqno (2.10a)
$$
$$
\{\psi _{-}(x),\bar \psi_{-}(y)\} =
-{i\over{2k}}(h_{11}+ h_{22})\delta ^{2}(x - y)\eqno(2.10b).
$$
As evident, new fields defined by (2.8)
are convenient only for $SU(2)$ and $SU(1,1)$  cases.  For $SL(2,R)$  case
brackets (2.10)  are  vanishing  and  more  convenient  to  use  other
variables. We can  rewrite the brackets in the compact form
$$
\{V_{i}(x),V_{j}(y)\} = - {16\pi\over k}\epsilon _{ij}\delta ^{2}(x - y),
\eqno(2.11a)
$$
$$
\{\psi _{+}(x),\bar\psi_{+}(y)\} = {i\over k}\kappa ^{2} \delta ^{2}(x - y)
,\eqno(2.11b)
$$
$$\{\psi _{-}(x), \bar\psi_{-}(y)\} = -{i\over k}\kappa ^{2}
\delta ^{2}(x - y), \eqno (2.11c)$$
where $\kappa ^{2} = + 1$ for $SU(2)$ and $\kappa ^{2} = - 1$ for $SU(1,1)$.
\par
The brackets (2.11) allow us interpret $V_{\mu }$ as an Abelian
CS field (the statistical field) and $\psi _{+}, \psi _{-}$
as  charged matter fields [22,24].
\par
The action in terms of this fields on the plane have the form
$$
S = \int dt\int d^{2}x
\{ -{k\over{32\pi}}\epsilon ^{\mu \nu \lambda }V_{\mu }
\partial _{\nu }V_{\lambda } + $$
$$
{ik\over 2}[(\psi _{+}\bar{D}_{0}\bar{\psi }_{+} -
\bar{\psi }_{+}D_{0}\psi _{+})
- (\psi _{-}\bar{D}_{0}\bar{\psi }_{-} -
\bar{\psi }_{-}D_{0}\psi _{-})]
$$
$$
- {k\over {2 \pi}}iq_{0}(\bar{D}_{-}\bar{\psi }_{+} -
\bar{D}_{+}\bar{\psi }_{-}) +
{k\over{2 \pi} }iq_{0}(D_{-}\psi _{+} -
D_{+}\psi _{-})\}
,
\eqno(2.12)$$
where $D_{\pm }= D_{1}\pm iD_{2}=
\partial _{\pm } - i/2 V_{\pm } , V_{\pm } = V_{1}\pm iV_{2} .
$
{}From (2.3) we recognize that the time components $V_{0}$ and $q_{0}$  of the
gauge
potential $J_{0}$  are  the  Lagrange  multipliers,  arbitrariness  of  which
guaranties the gauge invariance (covariance) of the topological action.
\par
Related constraints (2.3a) in
components $F = \sum^{3}_{a=1} F_{a}T_{a}$ gen erate $SU(2)$ or
$SU(1,1)$ algebra
$$
\{G_{a}({\bf x}),G_{b}({\bf y})\} =
c_{abc}G_{c}({\bf x})\delta ^{2}(x - y),
\eqno(2.13)$$
 where  rescaled constraints $G^{a}= - (ik/8\pi ) F^{a}$ have the form
$$
G_{+} = G_{1}+ iG_{2}= - {k\over{2 \pi} }(D_{-}\psi _{+} - D_{+}\psi _{-}),
$$
$$
G_{-} = G_{1}+ iG_{2}=
- {k\over{2 \pi }}(\bar{D}_{-}\bar{\psi }_{+} - \bar{D}_{+}\bar{\psi }_{-}),
\eqno(2.14)$$
$$
G_{3}= {k\over{8\pi}}[(\partial _{1}V_{2} -
\partial _{2}V_{1}) + 8\pi (|\psi _{+}|^{2} - |\psi _{-}|^{2})].
$$
The physical subspace of the TFT is defined by the constraint surface
$$G_{\pm} = 0, G_{0} = 0,$$
and any breaking of the topological symmetry  relates with a deviation
from this surface.
Constraints (2.14) form a part of the Euler-Lagrange equations for the action
(2.12):
$$
D_{0}\psi _{\pm } = {1\over{2\sqrt{\pi}}}D_{\pm }q_{0},\eqno(2.15a)
$$
$$
[D_{+}, D_{-}] = 8\pi \kappa ^{2}(|\psi _{+}|^{2} -
|\psi _{-}|^{2}),\eqno(2.15b)
$$
$$
[D_{0}, D_{\pm }] = -4\sqrt{\pi} \kappa ^{2}(\bar{q}_{0}\psi _{\pm } -
\bar{\psi }_{\pm }q_{0}),\eqno(2.15c)
$$
$$D_{+}\psi _{-}= D_{-}\psi _{+}.\eqno(2.15d)$$
\par
 Solution of this equations defines the mooving frame according to
$$
D_{0}{\bf n}_{+} = -2 q_{0} {\bf s}\qquad ,\eqno(2.16a)
$$
$$\partial _{0}{\bf s}
= q_{0} {\bf n}_{-} + \bar{q}_{0} {\bf n}_{+} ,\eqno(2.16b)$$
$$D_{\pm }{\bf n}_{+} = - 4\sqrt{\pi } \psi _{\pm }{\bf s} ,\eqno(2.16c)$$
$$
\partial _{\pm } {\bf s} = 2\sqrt{\pi}  (\psi _{\pm } {\bf n}_{-} +
\bar\psi_{\mp } {\bf n}_{+}), \eqno(2.16d)
$$
where fields $V_{0} , V_{\pm }$ and $q_{0} ,
\psi _{\pm }$ are given by relations
$$V_{0} = -i(\partial _{0} {\bf n}_{+} ,{\bf n}_{-}), V_{\pm } =
-i(\partial _{\pm } {\bf n}_{+},{\bf n}_{-}),\eqno(2.17a)$$
$$q_{0} = 1/2 (\partial _{0} {\bf s ,n}_{+}), \psi _{\pm }=
{1\over{4 \sqrt{\pi }}}(\partial _{\pm } {\bf s ,n}_{+}). \eqno(2.17b)$$

We note that the system (2.15), as well as (2.16), is invariant under
conformal transformations
$$
z' = f(z), \bar z' = \bar f(\bar z)$$
$$
V_{-} =  f^{\prime} V'_{-}, V_{+} = \bar f^{\prime} V'_{+},
$$
$$
\psi_{-} = f^{\prime} \psi'_{-}, \psi_{+} = \bar f^{\prime} \psi'_{+}.
$$
\par At the end of this section we reproduce some useful formulas
$$
(\partial _{\pm }{\bf s,\partial }_{\pm }{\bf s}) =
16\pi  \psi _{\pm }\bar\psi_{\mp },
\eqno(2.18)$$
$$
(\partial _{+}{\bf s, \partial }_{-}{\bf s}) =
8\pi  (|\psi _{+}|^{2} + |\psi _{-}|^{2}),
\eqno(2.19)$$
$$
\partial _{+}{\bf s \times  \partial }_{-}{\bf s} =
8i\pi  (|\psi _{+}|^{2} - |\psi _{-}|^{2}).
\eqno(2.20)$$
\medskip\bigskip
\noindent {\bf 3. Nonlinear} $\sigma $ - {\bf Model. Examples}
\bigskip
\par
In this section we describe some  simple  2  -  dimensional  models
in the formalism of Sec.1. The  first model is conformal  invariant,
while the second one is just integrable. In both cases time  evolution
is defined by the  Lagrange  multipliers $q_{0},V_{0}$ and  has
  an  arbitrary character.
Imposing  equations of motion for  the  model  as  constraints  on  the  field
variables we restrict the phase space of CS TFT.
\medskip
\par
\noindent 3.1.{\bf 2+0 dimensional $\sigma $ model}
\par
\medskip
As a first simple example we consider euclidean nonlinear $\sigma $  model
for the classical spin vector ${\bf s}$
$$\partial _{+}\partial _{-}{\bf s}
+ (\partial _{+}{\bf s}, \partial _{-}{\bf s}){\bf s} = 0.\eqno (3.1)$$
The model is  conformal  invariant.  This  fact  guarantees  that  the
conformal invariance of the CS TFT (2.12),   supplied with  eq.(3.1) will  be
preserved.
\par
Due to eqs.(2.16dc-d), (2.19) and  relation
\par
$$
\partial _{+}\partial _{-}{\bf s} =
2\sqrt{\pi}[(D_{-}\psi _{+}){\bf n}_{-}
+ (\bar{D}_{-}\bar{\psi }_{+}){\bf n}_{+})
- 8\pi  (|\psi _{+}|^{2} + |\psi _{-}|^{2}){\bf s}
\eqno(3.2)$$
the moving frame (2.16) and the field equations  (2.15), consistent
with eq.(3.1),  should
be supplied with additional constraints
$$D_{-}\psi _{+} = D_{+}\psi _{-} = 0 . \eqno(3.3)$$
\par
The resulting system (2.15) decouples on the evolutionary  part
$$
D_{0}\psi _{\pm } = {1\over{2\sqrt{\pi }}}D_{\pm }q_{0} ,\eqno (3.4a)
$$
$$
[D_{0}, D_{\pm }] = -4\sqrt{\pi} \kappa ^{2}(\bar{q}_{0}\psi _{\pm }
- \bar{\psi }_{\pm }q_{0}),\eqno (3.4b)
$$
which contains an arbitrary Lagrange multipliers $q_{0}, V_{0}$
and the spatial part
$$
D_{-}\psi _{+} = D_{+}\psi _{-} = 0,\eqno (3.5a)
$$
$$
[D_{+}, D_{-}] = 8\pi \kappa ^{2}(|\psi _{+}|^{2}
- |\psi _{-}|^{2}).\eqno (3.5b)
$$
The last system (3.5) is completely equivalent to the
$\sigma $ model (3.1)
and have a remarkable property. Most interesting for TFT may be that
eqs.(3.5),
known as the Hitchin equations,
 can be formulated on arbitrary Riemann surface.
\par
The system (3.5)(in contrast to eqs.(3.4))
 is invariant also under simple transformation
$$
\psi _{+} \rightarrow e^{i\gamma } \psi _{+} , \psi _{-} \rightarrow
e^{-i\gamma } \psi _{-} ,
$$
$$
V_{\pm } \rightarrow V_{\pm } ,
\eqno(3.6)$$
where $\gamma $ = constant.
This transformation for  $\sigma $ model is known as
the Pohlmeyer's $R^{(\gamma )}$ - transformation [29].
It  relates  to  a
"hidden" symmetry of the model and generates  an  infinite  set  of
non-local conservation laws. It seems not obvious as    this  symmetry
acts in the CS TFT (2.12).
\par If we attempt  to  describe  the  symmetry
transformation (3.6) as the global $U(1)$ gauge  transformation  (1.19),
(1.23),
we immediately obtain that one  of the
fields $\psi _{+} , \psi _{-}$ should
vanish.  As a result,  the   system   (3.5)   reduces   to   the
self-(anti)dual Chern-Simons equations [28]:
$$
D_{\pm } \psi _{\mp } = 0\qquad ,\eqno (3.7a)
$$
$$[D_{+}, D_{-}] = \pm  8\pi \kappa ^{2}|\psi _{\mp }|^{2},\eqno (3.7b)$$
related with  the  Liouville  equation.  The  instantons  (topological
solitons) of the  model  (3.1)   correspond  to  the  Chern-Simons
solitons of the model (3.7) [22], while
the topological charge (1.28) becomes of the  electric
charge form
$$
Q_{\pm } = \pm  \int |\psi _{\pm }|^{2}d^{2}x .
$$ \par
Solutions other than solitons, when  both $\psi _{+}$
and $\psi _{-}\neq  0,$ are  described
by  the  conformal  Sinh-Gordon  equation   [22]   (Toda   hierarchy)
reduced from(3.5). It is  worth  to  note  that  both  of  the systems
(3.5)  and   (3.7)   is   conformal and $R$  invariant. However,   only   the
self-dual system (3.7) admits the Pohlmeyer's symmetry (3.6),
as a gauge symmetry.  This  fact
intimately relates with the Darboux integrability  of  the  Liouville,
but not the Conformal Sinh-Gordon equation. Moreover, we  expect  that
it is connected also with special properties of the CS action,
admitting diffeomorphism invariance as gauge invariance.

\bigskip\par
\noindent 3.2 {\bf (1+1) - Heisenberg model}

\bigskip\par
If in the  previous  case  we  considered  the  conformal  invariant
integrable model, now we like to break the conformal invariance, but
preserve integrability. This is the well  known  classical  continuous
isotropic Heisenberg  model,
describing  precession  of  the  spin ${\bf s}$
according to the Landau-Lifshitz equation
$$
\partial _{0} {\bf s} =
{\bf s \times  \partial }_{j}\partial ^{j}{\bf s},
\eqno(3.8)$$
where ${\bf s}$ belongs to the 2-dimensional
sphere $S^{2}$ (or  pseudosphere
$S^{1,1}$).
\par
In this section we examine only  1+1  dimensional  case  (2+
1 dimensional model is considered in Sect.4).  We  will  treat  here  both
coordinates as the space coordinates. Then the model
\par$$
\partial _{1}{\bf s} = {\bf s \times  \partial }^{2}_{2}{\bf s}
\eqno(3.9)$$
is some kind of the 2-dimensional $\sigma $ model.
\par
Substituting
\par$$
\partial _{1}{\bf s} = q_{1} {\bf n}_{-}
+ \bar{q}_{1} {\bf n}_{+},\eqno(3.10a)
$$
$$
\partial ^{2}_{2}{\bf s} = D_{2}{\bf n}_{-}
+ \bar{D}_{2} {\bf n}_{+} - 4|q_{2}|^{2} {\bf s},\eqno (3.10b)
$$
to eq.(3.9) we find the constraint
between components
$$
q_{1}= iD_{2}q_{2} ,
\eqno(3.11)$$
where the covariant derivative $D_{2} = \partial _{2} - {i\over 2}V_{2}$.
Equation  (3.11)  allows
us exclude $q_{1}$ field from equations.
\par
The moving frame equations (1.22) now read
$$
D_{1}{\bf n}_{+} = - 2\hbox{iD}_{2}q_{2} {\bf s} ,\eqno (3.12a)
$$
$$
D_{2}{\bf n}_{+} = - 2 q_{2} {\bf s} ,\eqno (3.12b)
$$
$$\partial _{1}{\bf s} = iD_{2}q_{2} {\bf n}_{-} - i\bar D_{2}\bar q_{2}
{\bf n}_{+},
\eqno(3.12c)   $$
$$
\partial _{2}{\bf s} = q_{2} {\bf n}_{-} + \bar q_{2} {\bf n}_{+}
,\eqno (3.12d)$$
while the field equations are \par
$$
iD_{1}q_{2} + D^{2}_{2}q_{2} = 0,\eqno(3.13a)$$
$$
\partial _{1}V_{2} - \partial _{2}V_{1} =
 -4\kappa ^{2}\partial _{2}|q_{2}|^{2} .\eqno (3.13b)$$
In terms of redefined fields
$$A_{1} = V_{1} + 4\kappa ^{2}|q_{2}|^{2}, A_{2}= V_{2},\eqno (3.14)$$
(3.13) becomes
$$i(\partial _{1}- {i\over2} A_{1})q_{2}
+ (\partial _{2}- {i\over2} A_{2})^{2}q_{2}
- 2\kappa ^{2}|q_{2}|^{2}q_{2} = 0,\eqno (3.15a) $$
$$\partial _{1}A_{2} - \partial _{2}A_{1} = 0.\eqno (3.15b)$$
\par
The second equation allows one exclude the potentials
$A_{1}$ and $A_{2}$ by the $U(1)$ gauge transformation. If
$A_{j} = 2\partial _{j} \lambda $ we define new $\Phi  = q e^{i\lambda }$,
subject to  the  Nonlinear
Schr\"odinger equation
$$
i\partial _{1}\Phi  + \partial ^{2}_{2}\Phi
- 2\kappa ^{2}|\Phi |^{2}\Phi  = 0 .
\eqno(3.16)$$

As well known, this equation is integrable and  admits an infinite  number
of conservation laws, interpreted  as  continuity
equations in our case.
\par
The topological charge (1.28) appears now as
$$
Q = - \kappa ^{2}{2\over{\pi}}  \int d^{2}x \partial _{2}|\Phi |^{2}
= - \kappa ^{2}{2\over{\pi}}  \int dx_{1} (|\Phi (x_{1},x_{2}= L_{+})|^{2}
                                         - |\Phi (x_{1},x_{2}= L_{-})|^{2}),$$
where $L_{\pm}$ are the boundary values in the second space direction.
The usual evolution form for NLSE, when $x_{1} = t , x_{2} = x$ gives the
meaning of
$$
Q = {1\over{4\pi}}\int
{\bf s}\partial_{t} {\bf s}\times\partial_{x}{\bf s} dt dx
= - \kappa ^{2}{2\over{\pi }} \int_{x=C} dt |\Phi (x,t)|^{2}
$$
as a 1 dimensional Wess-Zumino  term.  It turns out that  well  known  soliton
solutions on infinite space line (the plane for $M_{2})$
and  periodic solutions
on the finite interval (the cylinder for $M_{2})$ always have vanishing Q.
Non-vanishing contribution should  appear  for  the  compact  on $(x,t)$
boundary condition (the Riemann surface for $M_{2})$.
\par
Worth  to  note  that  integrability  of  models (3.9)  and  (3.16)  is
connected with  the  Lax  pair  representation  or   ZCC  with  an
arbitrary spectral parameter. This loop  algebra  structure  provides
nonlocal conserved quantities generating non-Abelian algebra  for  the
NLSE [30]. They arise as a hidden non-Abelian structure of the model
(3.16). We understand now the
geometrical meaning of  this  structure,  since  the
model phase space can be considered  as  the  tangent  plane  to
2-dimensional  manifold,  being  sphere  for
$\kappa ^{2}= 1$   and
pseudosphere for $\kappa ^{2}= -1.$
Moreover, in the next  section  we  show  that
the spectral parameter has origin from the extra space direction.
\par
\bigskip \medskip
\noindent 4. {\bf Dimensional Reduction of 2+1 HM}
\par
\bigskip \medskip
{}From the action (2.3) or (2.12) we conclude that evolution of  the
model is completely defined by Lagrange multipliers. Usually,  to  fix
gauge freedom one uses the Hamiltonian gauge, when
\par
$$
q_{0} = 0, V_{0} = 0.
\eqno(4.1)$$
Thus, all considered field configurations  are  static.  In  this  case
reparametrization invariance of the theory corresponds to an arbitrary
time dependence for parameters of the moduli space.  But  if  we  like
study an integrable deformations of the topological symmetry  we  need
consider  more  restricted  gauge  conditions,  providing   integrable
dynamics.
\par
In  the  present  section  we  like  to  choose  a   different   gauge
condition. Since evolution equation for $\sigma $ models  in  tangent  space
reduces to constraints on the phase space variables we can choose this
constraints as a new gauge conditions. Then, the resulting CS theory
should have the corresponding time evolution [22-24].
\par
We consider the HM  (3.8) in 2+1 dimensions,
$$
\partial _{0} {\bf s} = {\bf s \times } (\partial ^{2}_{1} +
\partial ^{2}_{2} ){\bf s},
\eqno(4.1)$$
with $({\bf s},{\bf s}) = 1.$
\par
{}From the mooving frame equations  (1.22)  we  conclude  that  eq.(4.1)
leads to the constraint on $q_{0}$ :
$$
q_{0}= iD_{1} q_{1} + iD_{2} q_{2}
\eqno(4.2)$$
This relation allows one exclude $q_{0}$ from the system (1.22) and we have
$$
D_{0}{\bf n}_{+} = - 2i(D_{k}q_{k}) {\bf s} ,\eqno (4.3a)
$$
$$
D_{k}{\bf n}_{+} = - 2 q_{k} {\bf s} ,\eqno (4.3b)
$$
$$
\partial _{0}{\bf s} = i(D_{k} q_{k} ){\bf n}_{-} -
i (\bar D_{k}\bar q_{k} ){\bf n}_{+} ,\eqno (4.3c)$$
$$
\partial _{k} {\bf s} = q_{k} {\bf n}_{-} + \bar q_{k} {\bf n}_{+} ,
(k = 1,2) .
\eqno (4.3d)$$
Remaining field variables should satisfy to the system
$$
iD_{0}q_{k} + D_{k}D_{l}q_{l} = 0 ,\eqno (4.4a)
$$
$$
D_{k}q_{l} = D_{l} q_{k} , (k,l = 1,2), \eqno(4.4b)
$$
$$
[D_{k}, D_{l}] = -2\kappa ^{2}(\bar q_{k}q_{l} - q_{k}\bar q_{l}) ,\eqno (4.4c)
$$
$$
[D_{0}, D_{k}] = 2i\kappa ^{2}(q_{k}\bar D_{l}\bar q_{l}
+ \bar q_{k} D_{l} q_{l}) .\eqno (4.4d)
$$
In terms of complex fields (2.8)
$$
\psi _{\pm } = {1\over{2\sqrt{\pi}}}  (q_{1}\pm iq_{2}) ,
\eqno(4.5)$$
we have for the moving frame
$$
D_{0}{\bf n}_{+} = - 4i\sqrt{\pi}  (D_{-}\psi _{+}) {\bf s} +
 4i\pi \kappa ^{2} (|\psi _{+}|^{2}
+ |\psi _{-}|^{2}){\bf n}_{+} ,\eqno (4.6a)
$$
$$
D_{\pm }{\bf n}_{+} = - 4\sqrt{\pi}  \psi _{\pm } {\bf s} ,\eqno (4.6b)
$$
$$\partial _{0}{\bf s} = 2i\sqrt{\pi} ((D_{-}\psi _{+} ){\bf n}_{-}
- (\bar{D}_{-}\bar{\psi }_{+}){\bf n}_{+}) ,\eqno (4.6c) $$
$$
\partial _{\pm }{\bf s} = 2\sqrt{\pi} (\psi _{\pm } {\bf n}_{-}
+ \bar\psi_{\mp } {\bf n}_{+})\eqno (4.6d)$$
and for the field equations
\par
$$
iD_{0}\psi _{\pm } + (D^{2}_{1}+ D^{2}_{2})\psi _{\pm }
+ 8\pi \kappa ^{2}|\psi _{\pm }|^{2}\psi _{\pm } = 0 , \eqno(4.7a)$$
$$
D_{-}\psi _{+} = D_{+}\psi _{-}\eqno (4.7b) $$
$$
[D_{+}, D_{-}] = 8\pi \kappa ^{2}(|\psi _{+}|^{2}
- |\psi _{-}|^{2}),\eqno (4.7c)$$
$$
[D_{0}, D_{\pm }] = 8i\pi \kappa ^{2}(\bar\psi_{\mp }D_{\pm } \psi _{\mp }
+ \psi _{\pm }\bar D_{\mp }\bar \psi_{\pm }) -
4i\pi \kappa ^{2} \partial _{\pm }(|\psi _{+}|^{2}
+ |\psi _{-}|^{2}).\eqno(4.7d) $$
The covariant derivatives in the system (4.6) and (4.7) are
defined as before (see eq.(2.12)):
$$
D_{\pm } = \partial _{\pm } - {i\over2} A_{\pm } ,
D_{0} = \partial _{0} - {i\over2} A_{0},
$$
with $A_{\pm } = V_{\pm }$ , but redefined
$$
A_{0} = V_{0} - 8\pi \kappa ^{2}(|\psi _{+}|^{2} + |\psi _{-}|^{2}).
$$
\par For the static field configurations, eq.(4.1) reduces to the $\sigma$
model (3.1), considered before.
Now we are going to perform a dimensional reduction of the  model
to have an integrable evolution system. Let $M = T \times  R \times  S$
,  where $R$ is real line associated
with the $x_{1}$ space coordinate and $S$  is
 compactified on the circle second space coordinate $x_{2}$.
As usual, for zero modes we are looking for equations independent of
$x_{2},$
\par
$$
iD_{0}\psi _{\pm } + [D^{2}_{1}+ ({i\over2} A_{2})^{2}]\psi _{\pm }
+ 8\pi \kappa ^{2}|\psi _{\pm }|^{2}\psi _{\pm } = 0 , \eqno(4.10a)$$
$$
(\partial _{1} - {i\over2} A_{-}) \psi _{+} = (\partial _{1} - {i\over2} A_{+})
\psi _{-},\eqno (4.10b)$$
$$
\partial _{1}A_{2} = -8\pi \kappa ^{2}(|\psi _{+}|^{2}
- |\psi _{-}|^{2}),\eqno (4.10c)$$
$$
\partial _{0}A_{1} - \partial _{1}A_{0}
= 8\pi \kappa ^{2}A_{2} (|\psi _{+}|^{2} - |\psi _{-}|^{2}),\eqno(4.10d)$$
$$
\partial _{0}A_{2} = -8i\pi \kappa ^{2}[(\bar\psi_{+}
\partial _{1} \psi _{+} - \psi _{+} \partial _{1}\bar\psi_{+} )
- (\bar\psi_{-} \partial _{1} \psi _{-}
- \psi _{-} \partial _{1}\bar\psi_{-} )
$$
$$- iA_{1}(|\psi _{+}|^{2} - |\psi _{-}|^{2})].\eqno (4.10e)$$
If instead of the potential $A_{0}$ we introduce
\par
$$
{\cal A}_{0} = A_{0} - {1\over2} A^{2}_{2} , {\cal A}_{1}= A_{1} ,
{\cal A}_{2}= A_{2},
\eqno(4.11)$$
\noindent equation (4.10d) becomes of the vanishing field strength
form \par
$$
\partial _{0}{\cal A}_{1} - \partial _{1}{\cal A}_{0} = 0,\eqno (4.12d)$$
\noindent and for the rest equations we have
\par
$$
i{\cal D}_{0}\psi _{\pm } + {\cal D}^{2}_{1}\psi _{\pm }
+ 8\pi \kappa ^{2}|\psi _{\pm }|^{2}\psi _{\pm } = 0 , \eqno(4.12a)$$
$$
(\partial _{1} - {i\over2} {\cal A}_{-}) \psi _{+}
= (\partial _{1} - {i\over2} {\cal A}_{+})\psi _{-},\eqno (4.12b)$$
$$
\partial _{1}{\cal A}_{2} = -8\pi \kappa ^{2}(|\psi _{+}|^{2}
- |\psi _{-}|^{2}),\eqno (4.12c)$$
$$\partial _{0}{\cal A}_{2} = -8i\pi \kappa ^{2}
[(\bar\psi_{+}{\cal D}_{1} \psi _{+} - \psi _{+} \bar{\cal D}_{1}\bar
\psi_{+} )
- (\bar\psi_{-}{\cal D}_{1} \psi _{-}
- \psi _{-}\bar{\cal D}_{1}\bar\psi_{-} ). \eqno(4.12e)$$
Comparing  eqs.(4.12a)  and (4.12d)  with  gauged  NLSE (3.15a-b) we
recognize complete equivalence. Using the same as before procedure, we  can
compensate
the gauge potentials via $U(1)$ rotation
\par
$$
{\cal A}_{0} = 2\partial _{0} \lambda  , {\cal A}_{1}
= 2\partial _{1} \lambda  , \Phi _{\pm } = \psi _{\pm } e^{i\lambda}
.\eqno(4.13)$$
\noindent Thus,
 we find that both of the $\Phi _{+}, \Phi _{-}$
fields is  satisfy to the NLSE
\par
\noindent
$$i\partial _{0}\Phi _{\pm } + \partial ^{2}_{1}\Phi _{\pm }
+ 8\pi \kappa ^{2}|\Phi _{\pm }|^{2}\Phi _{\pm } = 0 , \eqno(4.14)$$
\par
\noindent and to the set of relations connecting
${\cal A}_{2}$ with $\Phi _{+}, \Phi _{-}$ fields
\par
$$
\partial _{1}\Phi _{+} - \partial _{1}\Phi _{-}
= 1/2 {\cal A}_{2}(\Phi _{+} + \Phi _{-}),\eqno (4.15a)$$
$$
\partial _{1}{\cal A}_{2} = -8\pi \kappa ^{2}(|\Phi _{+}|^{2}
- |\Phi _{-}|^{2}),\eqno (4.15b)$$
$$
\partial _{0}{\cal A}_{2} =
-8i\pi \kappa ^{2}[(\bar\Phi_{+} \partial _{1} \Phi _{+}
- \Phi _{+} \partial _{1}\bar\Phi_{+} ) -
(\bar\Phi_{-} \partial _{1} \Phi _{-} -
\Phi _{-} \partial _{1}\bar\Phi_{-})]. \eqno(4.15c)$$
\noindent The system (4.15) allows one define the ${\cal A}_{2}$ field
in explicit form
\par

$${\cal A}_{2} = \epsilon _{\pm }\sqrt{ \alpha ^{2}_{0} -
16\pi \kappa ^{2}|\Phi _{+}- \Phi _{-}|^{2}},\eqno (4.16)$$
where $\alpha _{0}$  is  the  integration  constant,
$\epsilon _{\pm } = \pm 1.$
I can easily
show  that if both $\Phi _{+}$
and $\Phi _{-}$ are solutions of the  NLSE  (4.14),  then  the evolution for
${\cal A}_{2}$ (4.16) is satisfied to eq.(4.15c).
\par
But eq.(4.15a) with (4.16) are just the  B\"acklund transformations  for
the NLSE (4.14). Thus, the surprising moment arising from (2+1) dimensional
reduction is an interpretation of the B\"acklund transformation
for NLSE in terms of the  Abelian  Chern-Simons  gauge  field,
 associated with the extra space coordinate $x_{2}$.
\par
When $\Phi _{+}= \Phi _{-}$ the ${\cal A}_{2} =
\epsilon _{\pm }\alpha _{0} =  const.$  As  we  show  immediately  this
constant have  meaning of the spectral parameter.
When $\Phi _{+} \neq  \Phi _{-}$,
that means a soliton creation, ${\cal A}_{2}$
is inhomogeneuos function  measuring
the departure of $\Phi _{+}$ from $\Phi _{-}$.
\par
To clarify the meaning of the homogeneous part
$\alpha _{0}$ we turn  now  to  the
chiral current (1.7a)
\par
$$
J _{\mu} = g^{-1} \partial _{\mu} g = {i\over4}\sigma_{3}V_{\mu}
+ \left(\matrix{0&-\kappa^{2}\bar q_{\mu} \cr q_{\mu}&0 \cr }\right).
\eqno (4.17)$$
\noindent We  can  carry out
the $U(1)$   gauge
transformation
\par
$$
g \rightarrow g e^{{i\over4}\lambda  \sigma _{3}}.
\eqno(4.18)$$
\noindent As a result we have
\par
\noindent $$
J_{1} = \sqrt{\pi} \left(\matrix{0& -\kappa^{2}(\bar\Phi_{+} + \bar\Phi_{-})
\cr
\Phi_{+} + \Phi_{-} & 0\cr}\right),
$$
\par
$$
J_{2} = {i\over4}\sigma_{3}{\cal A}_{2} - i\sqrt{\pi}
\left(\matrix{0&\kappa^{2}(\bar\Phi_{+} - \bar\Phi_{-})\cr \Phi_{+} -
\Phi_{-}&0\cr}\right),
$$
$$
J_{0} = {i\over8}\sigma_{3}[{\cal A}^{2}_{2} + 16\pi\kappa^{2}
(|\Phi_{+}|^{2} + |\Phi_{-}|^{2})] + i\sqrt{\pi}
\left(\matrix{0&\kappa^{2}
(\bar{\cal D}_{-}\bar\Phi_{+} + \bar{\cal D}_{+}\bar\Phi_{-}\cr
{\cal D}_{-}\Phi_{+} + {\cal D}_{+}\Phi_{-}& 0\cr }\right).
$$
\noindent Using (4.7b)
$$
{\cal D}_{-}\Phi_{+} = {\cal D}_{+}\Phi_{-} ,
$$we can rewrite it as
\par
$$
J_{+} = - {1\over4}\sigma_{3}{\cal A}_{2} + 2\sqrt{\pi}
\left(\matrix{0  & -\kappa^{2}\bar\Phi_{-}\cr \Phi_{+} & 0\cr  }\right),
\eqno (4.19a)$$
$$
J_{0} = {i\over8}\sigma_{3}[{\cal A}^{2}_{2} + 16\pi\kappa^{2}
(|\Phi_{+}|^{2} + |\Phi_{-}|^{2})] + 2i\sqrt{\pi}
\left(\matrix{ 0 & \kappa^{2}\bar{\cal D}_{+}\bar\Phi_{-}  \cr
{\cal D}_{-}\Phi_{+}      &    0   \cr  }\right).
\eqno (4.19b)$$
\par
Now, let
\par
$$
\Phi _{+} = \Phi _{-} \equiv  {1\over{2\sqrt{\pi}}}\Phi
,\eqno (4.20)$$
\noindent then, from (4.16)
\par
$$
{\cal A}_{2} = \epsilon _{\pm }\alpha _{0} \equiv  4\lambda _{0}
.\eqno (4.21)$$
\noindent As a result we have the well known Lax pair for NLSE
\par
$$
J_{+} = - \lambda _{0}\sigma_{3} + \left(
\matrix{0  & -\kappa^{2}\bar\Phi  \cr
\Phi  & 0 \cr }\right),
\eqno (4.22a)$$
$$
J_{0} = i\sigma_{3}[2\lambda^{2}_{0} + \kappa^{2}|\Phi|^{2}] + i
\left( \matrix{ 0 & \kappa^{2}(\partial_{1} + 2\lambda_{0})\bar\Phi   \cr
(\partial_{1} - 2\lambda_{0})\Phi    &    0  \cr  }\right).
\eqno (2.22b)$$
The Lax pair for the 1+1 HM model (3.9) can be constructed from (4.22)  by
usual
procedure of the gauge transformation, in terms of (1.2) trihedral $N_{3}$
[31,32].
\noindent
\par It is clear now, that  constant
$\lambda _{0}$ has a  meaning  of  the  spectral
parameter.
 Remarkable fact is that $J_{+}$ consists of
two parts: $J_{1}$ part is
independent of the spectral parameter and $J_{2}$ is completely
defined in terms of it.
\par
As known, in order to  investigate  the  infrared  properties  of  the
theory, we can expand the  gauge
field ${\cal A}_{2}$ in  a  Fourier  series  and
separate out the part which plays main role at long distances.
This is the constant  in
space ($x_{1}$) term
\par
$$
{\cal A}_{2}\cong  4\lambda _{0}(1
+ {{\pi \kappa ^{2}}\over{2\lambda ^{2}_{0}}}|\Phi _{+}- \Phi _{-}|^{2}
+ \ldots).
\eqno (4.23)$$
\noindent Thus we can interpret the spectral
parameter as a condensat value  for
the  Chern-Simons  gauge  field   ${\cal A}_{2}$
associated  with  the   extra
dimension.
\par
As we see,   the Lax pair with the
spectral parameter flow , defining all the miracles of soliton mathematics,
has  a
simple interpretation in terms of an  extra space direction and CS TFT.
\vfill\eject
\noindent
{\bf Conclusion}
\bigskip
\par
In conclusion I like to emphasize some points.
First, as shown above, the non-Abelian TFT (2.1) represented in the form
(2.12)
could be interpreted as the Abelian Chern-Simons gauge theory interacting with
a doublet of matter fields [24]. Usually, the Abelian gauge field is called
the "statistical
field" since  defines the anyonic statistics for matter fields.
For more direct relation we need rescale the "matter" fields
$$\psi_{\pm} = {1\over {\sqrt k}} \Psi _{\pm},$$
to have normal canonical brackets (2.11). Then, the Chern-Simons Gauss law
(2.15b) becomes
$$\partial_{1}V_{2} - \partial_{2}V_{1} = - {{8 \pi \kappa^{2}}
\over k}(|\Psi_{+}|^{2}
- |\Psi_{-}|^{2}).$$
{}From this form we recognize that the coupling constant $k$ for non-Abelian
theory (2.1) coincides with the statistical parameter for fractional
statistics. It means that in quantized (2.1) theory the mater fields
could appear (after singular gauge transformation) as anyons [24].
\par
The 1+1 dimensional reduction of the model (4.7) shows that two components
${\cal A}_{0}, {\cal A}_{1}$ of the statistical gauge field can be removed by
gauge
transformation. But component ${\cal A}_{2}$ (4.16 ) related to the extra space
coordinate has a deep physical meaning. Thus, for infrared
properties of the Chern-Simons theory (2.12) only the constant in space vector
potential
$$A_{i}(x) = X_{i}(t) + ... $$
depending only of time, is  relevant. Corresponding Chern-Simons term in (2.12)
$${k\over{16\pi}}\dot{\bf X}\times {\bf X} ,$$
has a simple physical interpretation.
If we consider $(X_{1},X_{2})$ as
coordinates of the charged particle in the plane, and switch on the magnetic
field
orthogonal to the plane, the Lorentz force will arise. It connects two
directions
$X_{1}, X_{2}$ in such a way that energy from the first direction will flow
to the second one.
In our case it means that due to the Chern-Simons structure in the topological
action (2.12), our 1+1 dimensional gauge theory
(4.14),(4.15) continue to feel an extra space
coordinate. But all dependence of the extra space coordinate is hidden in
the spectral parameter (4.21). Of course the gauge invariant nonlinear
equations
(4.14) are independent of $x_{2}$ and $\lambda_{0}$.
\par
Thus, the potential (4.16) generally includes two parts.
 Part with the  spectral
parameter is a constant, and has the meaning of the condensate for
statistical gauge field.  While non-homogeneous part of ${\cal A}_{2}$
comes from the deviation between two solutions of the NLSE (4.14). We know
that (4.15-16) is an elementary B\"acklund transformation for NLSE.
If one of the
fields is vanishing, it provides the one-soliton solution for the second one.
Corresponding value of ${\cal A}_{2}$ we call the one-soliton gauge potential.
This allows us to formulate a gauge invariance principle. The
statistical gauge field  is homogeneous, global defined field in the
case (4.20). But when condition (4.20) is broken, that means a soliton is
created, the gauge field ${\cal A}_{2}$ becomes a local function of coordinat
$x_{1}$. Hence, we observe that statistical gauge field is inseparable
phenomena accompanying the soliton creation. It is a relict of the Chern-Simons
Gauss law which states the creation of a magnetic flux by particle creation.
In anyon physics we interpret the physical excitations
as particles with attached magnetic flux. In this sense we can
interpret our result in the next way. Even the one-dimensional solitons
 are excitations attached
with statistical magnetic field. Indeed, if we put one of the fields
, say $\Phi_{-} = 0$, from eq.(4.15b) follows the one-dimensional (!)
CS Gauss law
$$
B(x) = {{8\pi\kappa^{2}}\over k}|\Psi_{+}|^{2} .$$
For soliton with large amplitude $\eta = Im \lambda_{0}$,
we can write approximately
$$B(x) = {{16\pi\kappa^{2}}\over k}\eta\delta (x).$$
This relation should be compared with 2-dimensional "pro-totype" (1.36).
It shows explicitly that one soliton is always attached with "magnetic" field.
The line integral (one-dimensional flux)
$$
\int B dx =
{{8\pi\kappa^{2}}\over k} \int |\Psi_{+}|^{2} dx =
{{16\pi\kappa^{2}}\over k}\eta,
$$
is time independent and well known first integral of NLSE. It has a
simple interpretation of the rescaled soliton amplitude $Im \lambda_{0}$
and really is inseparable from soliton in any collisions.

\bigskip
\noindent
{\bf Acknowledgments}
\bigskip
\medskip
\par

The author would like to thank Professor Loriano Bonora,
for kind hospitality at the International School for Advanced Studies
(SISSA/ISAS), Trieste.\par

He would also like to thank Professor Loriano Bonora and Professor
Boris Dubrovin for useful
discussions. This work was supported by SISSA contract N 5404.
\medskip \par
\vfill\eject

\noindent
{\bf References}
\medskip
\bigskip
\par \noindent
[1] T.Appelquist,A.Chodos and P.G.O.Fruend, {\it Modern Kaluza-Klein
Theories} Addison-Wesley Pub. (1987)
\par \noindent
[2] M Green J Schwarz E Witten {\it Superstring Theory} Camb.Univ.Press (1987)
\par \noindent
[3] N.S.Manton, Nucl.Phys. {\bf 158} 141 (1979)
\par \noindent
[4] J.H.Schwarz, Nucl.Phys. {\bf B 447} 137 (1995)
\par \noindent
[5] L.Dolan, Phys.Rep. {\bf 109} 1 (1984)
\par \noindent
\par \noindent
[6] R S Ward Phil.Trans.Roy.Soc.London {\bf A315} 451 (1985);
M Ablowitz , S Chakravarty and L A Takhtajan , Comm.Math.Phys. {\bf 158}
289 (1993)
\par \noindent
[7] A A Belavin and V.E.Zakharov, Phys.Lett. {\bf B73} 53 (1978)
\par \noindent
[8] L.D.Faddeev, Schladming lectures, 1995
\par \noindent
[9] A.B.Zamolodchikov, Advanced Studies in Pure Mathematics {\bf 19} 642 (1989)
\par \noindent
[10] F.Calogero, in {\it What Is Integrability?} ed. V E Zakharov Springer
(1991)
\par \noindent
[11] O.Babelon and L.Bonora, Phys.Lett.{\bf B244} 220 (1990)
\par \noindent
[12] V.Sokolov, private communication
\par \noindent
[13] E.Witten, Comm.Math.Phys. {\bf 117} 353 (1988)
\par \noindent
[14] D.Birmingham, M.Blau, M.Rakowski and G.Thompson, Phys.Rep. {\bf 209} 129
(1991)
\par \noindent
[15] L.Baulieu, Phys.Lett. {\bf B232}  473  (1989)
\par \noindent
[16] E.Witten, Comm.Math.Phys. {\bf 121} 351  (1989)
\par \noindent
[17] S.Elitzur, G.Moore, A.Schwimmer and N Seiberg , Nucl Phys {\bf B326}
108 (1989)
\par \noindent
[18] M.Blau and G.Thompson, Ann. Phys. {\bf 205} 130 (1991)
\par \noindent
[19] E.D'Hoker, Phys.Lett. {\bf B264} 101 (1991); K.Aoki and E.D'Hoker, Nucl.
Phys. {\bf 387}  567 (1992)
\par \noindent
[20] D.Cangemi, Phys.Lett. {\bf B297}  261 (1992) ; A.Achucarro, Phys.Rev.Lett.
{\bf 70}  1037 (1993) and Ref.
\par \noindent
[21] A.D'Adda,M.L\"uscher and P.Di Vecchia, Phys.Rep. {\bf 49}  239 (1979)
\par \noindent
[22] L.Martina, O.K.Pashaev and G.Soliani, Mod. Phys. Lett. {\bf A8}, 3241;
Phys.Rev. {\bf B48} 15 787 (1993)
\par \noindent
[23] O.K.Pashaev, {\it Integrable Chern-Simons Gauge Field Theory in 2+1
Dimensions},
ICTP Report IC/95/53, hep-th /9505178, Mod.Phys.Lett. A (to be published)
\par \noindent
[24] L.Martina, O.K.Pashaev and G.Soliani, {\it Topological Field Theory and
Nonlinear
$\sigma-$ models on Symmetric Spaces} , Lecce Univ. preprint DFUL - 1/06/95,
hep-th /9506130
\par \noindent
[25] H.Verlinde, Nucl.Phys. {\bf B 337}  652 (1990) ;
A.Bilal, V.Fock and I.Kogan, Nucl.Phys. {\bf 359}  635 (1991) ;
 A.Bilal, Phys.Lett. {\bf B 267}
 487 (1991)
\par \noindent
[26] S.Orfanidis, Phys.Rev. {\bf D21} 1513 (1980)
\par \noindent
[27] G.Thompson, {\it 1992 Trieste Lectures on Topological Gauge Theory and
Yang-
Mills Theory}, ICTP series vol.9, eds: E Gava et al,  World Sci (1993)
\par\noindent
[28] R.Jackiw, S.Y.Pi, Phys.Rev.Lett. {\bf 66} 2682 (1991); Phys.Rev. {\bf D42}
3500 (1990)
\par\noindent
[29] K.Pohlmeyer, Comm.Math.Phys. {\bf 46} 207 (1976);
H.Eichenherr and K.Pohlmeyer, Lett.Math.Phys. {\bf 2} 181 (1978)
\par \noindent
[30] H.Eichenherr, in {\it Group Theoretical Methods in Physics} Lect. Notes
in Phys. {\bf 180} 91 (1983)
\par\noindent
[31] V.E. Zakharov and L.A.Takhtajan, Teor.Math.Phys. {\bf 38} 26 (1979)
\par\noindent
[32] V.Makhankov and O.K.Pashaev, Soviet Sci.Rev./sect.C, {\bf 9} part3
 (1992)\end